\documentclass[multphys,vecphys]{svmult}

\usepackage{makeidx}         
\usepackage{graphicx}        
\usepackage{multicol}        
\usepackage[bottom]{footmisc}

\makeindex             

\begin{document}

\title*{Spectroscopic companions of very young brown 
dwarfs}
\author{Joergens Viki\inst{1} 
}
\institute{Leiden Observatory,
           Netherlands,
\texttt{viki@strw.leidenuniv.nl}\\
\textit{Proceeding of ESO workshop 'Multiple Stars across the HRD', 
Garching, 2005.}
}
%
%
\maketitle

\begin{abstract}
I review here the results of the first RV survey for spectroscopic companions
to very young brown dwarfs (BDs) and (very) low-mass stars in the ChaI
star-forming cloud with UVES at the VLT.
This survey studies the binary fraction in an as yet unexplored
domain not only in terms of primary masses (substellar regime)
and ages (a few Myr) but also in terms of companion masses
(sensitive down to planetary masses) and separations ($<$ 1\,AU). 
The UVES spectra obtained so far hint at spectroscopic
companions of a few Jupiter masses around one BD and around
one low-mass star (M4.5) with orbital periods of at least several months.
Furthermore, the data indicate a multiplicity fraction
consistent with field BDs and stellar binaries for periods $<$100 days.
\end{abstract}

\section{Introduction}

The multiplicity properties of brown dwarfs (BDs) are key parameters for
their formation. For example, embryo-ejection scenarios
predict few binaries in only close orbits (see Delgado-Donate, this
proceeding \cite{joe:delgado}), while isolated
fragmentation scenarios allow for an abundance of binaries over a wide
range of separations.  
\begin{figure}[b]
\centering
\includegraphics[height=0.7\textwidth,angle=0]{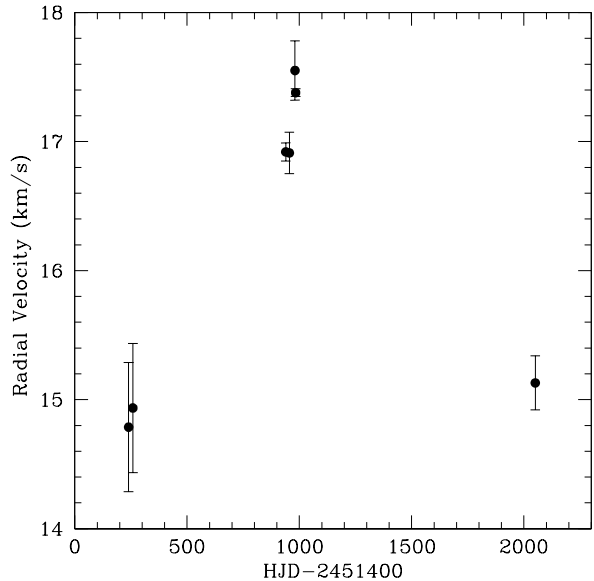}
\caption{
\label{joe:fig2}
Radial velocity data for the young BD candidate
Cha\,H$\alpha$\,8 (M6.5) recorded with UVES/VLT: significant
variability occurring on time scales of months to years hint at a companion
at $a>0.2\,$\,AU and a mass 
Msin\,$i$ of at least 6 M$_\mathrm{Jup}$ 
(Joergens 2005 \cite{joe:j2005}).
}
\end{figure}
In recent years, numerous low-mass and BD binaries were detected by direct
imaging in the field and in young clusters and associations
(see Bouy, this proceeding \cite{joe:bouy}).
Based on these observations, it was found that
very low-mass stars (VLMSs) and BDs pair less frequently in binary systems
than solar-like stars. However, these surveys cannot resolve the inner
$\sim$3 to 10\,AU (depending on distance) around the objects.
Companions orbiting at such close separations may have originated
based on a substantially different companion formation mechanism than
the one found so far by direct imaging. They can be detected
indirectly by spectroscopic Doppler surveys.
Precise monitoring of the radial velocities (RVs) of BDs
became possible in the last years with the generation of 8--10\,m class telescopes.
Shortly after the high-resolution echelle spectrograph UVES at the VLT
saw its first light in October 1999, two programs were started with this instrument
in spring 2000 and 2001
in order to systematically search for close companions to BDs and VLMSs
in the very young
Cha\,I star-forming region (Joergens \& Guenther 2001 \cite{joe:j2001};
Joergens 2005 \cite{joe:j2005}) and in the field (Guenther \& Wuchterl 2003 \cite{joe:gw}).
In this article, we present and discuss the current results of the survey in Cha\,I.

\section{Results of RV survey in Cha\,I}

In order to probe BD multiplicity at very young ages an
RV survey of BDs/VLMSs in the Cha\,I star-forming region ($\sim$2\,Myr) was started
by Joergens \& Guenther (2001 \cite{joe:j2001}).
Among a subsample of ten BDs/VLMSs (M$\leq$0.12\,M$_{\odot}$, M5--M8),
nine do not show signs of companions down to masses of giant planets.
The sampled orbital periods for them are $<$ 40 day corresponding to
separations $<$0.1\,AU (Joergens 2005 \cite{joe:j2005}).
Monte-Carlo simulations show that the data set
provides a fair chance ($>$ 10\%) to detect companions even out to 0.4\,AU (Maxted \& Jeffries 2005
\cite{joe:maxted}).
For the BD candidate Cha\,H$\alpha$\,8 (M6.5), RV data were recorded with a time base
of a few years and they indicate the existence of a spectroscopic companion of planetary or BD mass
with an orbital period of at least several months or even years
(Fig.\,\ref{joe:fig2}).
Thus, among ten BDs/VLMSs, none shows signs of BD companions
with periods less than 40 days and one with a period $>$ 100 days.
Furthermore, the low-mass star CHXR74 (M4.5)
also exhibit long-term RV variations (Fig.\,\ref{joe:fig3}) that are 
attributed to an orbiting companion
with at least 19 Jupiter masses.

\newpage
\section{Detection of RV planets around BDs in Cha\,I feasible}

\begin{figure}[b]
\centering
\includegraphics[height=\textwidth,angle=90]{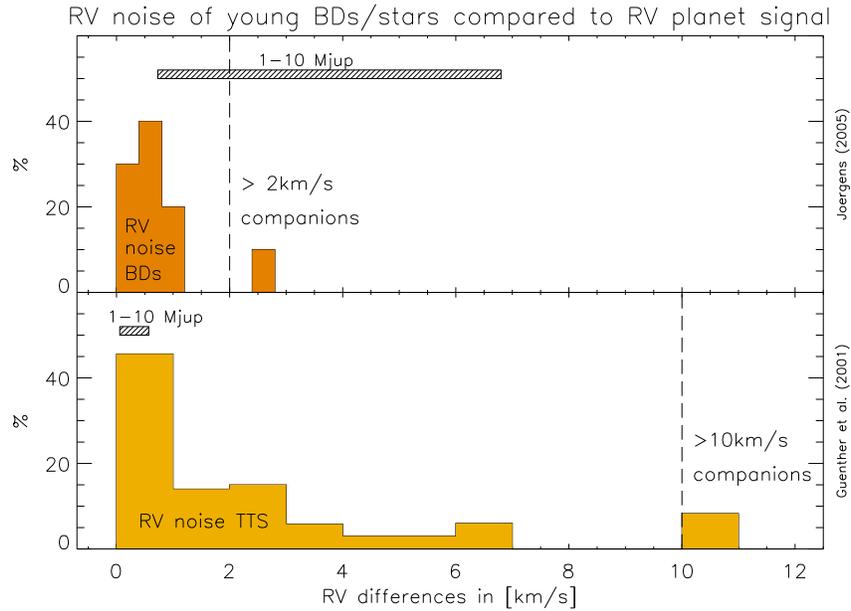}
\caption{
\label{joe:fig1}
Distribution of RV differences for very young BDs/VLMSs (upper panel, Joergens
2005 \cite{joe:j2005})
and for T~Tauri stars (lower panel, Guenther et al. 2001 \cite{joe:guenther}).
The plot shows that systematic errors caused by activity at very young ages are sufficiently
small in the
substellar mass regime to allow for detections
of giant planets by the RV technique, which is not the case for T~Tauri stars.
RV differences/amplitudes in the plot are peak-to-peak values.
Inserted ranges of RV planet signals were calculated based on circular orbits,
a semi major axis of 0.1\,AU and a primary mass of 0.06\,M$_{\odot}$ for BDs, and
1\,AU and 1\,M$_{\odot}$ for T~Tauri stars, resp.
}
\end{figure}

The RV survey in Cha\,I revealed that very young BDs/VLMSs
exhibit no RV noise due to surface activity down to the precision required to detect
Jupiter mass planets (Joergens 2005 \cite{joe:j2005}). 
They are, therefore, suitable targets when using the RV technique to search for planets.
In the upper panel of Fig.\,\ref{joe:fig1}, the distribution of RV differences recorded
for very young BDs and VLMSs (M$\leq$0.12\,M$_{\odot}$, M5--M8)
is compared to the RV signal caused by a planet
of 1-10 M$_\mathrm{Jup}$. It shows that the RV signal
of a giant planet is detectable well above the RV noise level for a BD primary.
The lower panel of Fig.\,\ref{joe:fig1} displays as comparison RV data
for T~Tauri stars recorded by Guenther et al. (2001 \cite{joe:guenther}).
The RV amplitude of a planet is completely swallowed by
the large systematic RV errors of up to
7\,km/s for them making planet detections by the RV technique around very young
stars quasi impossible.

\clearpage
\newpage

\section{Is the BD desert a scalable phenomenon?}

\begin{figure}[h]
\centering
\includegraphics[width=\textwidth,angle=0]{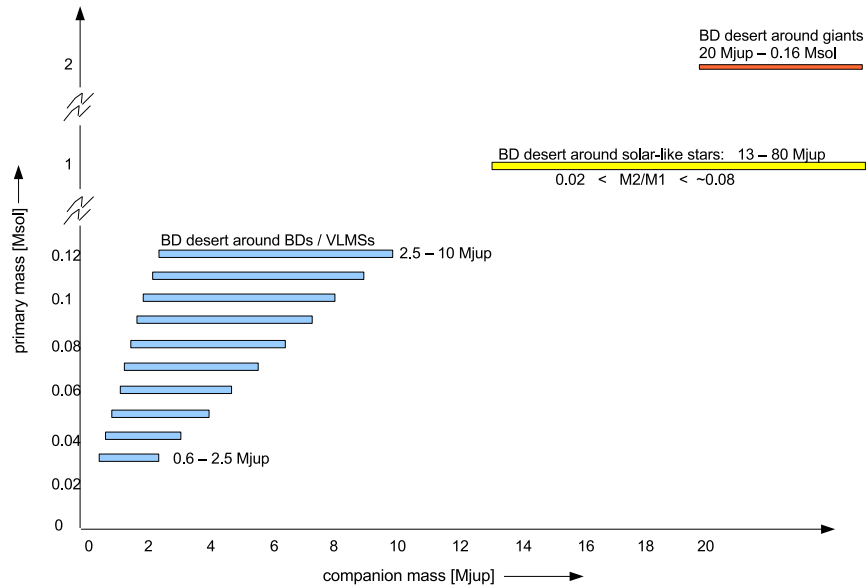}
\caption{
\label{joe:fig4}
Schematic illustration of the BD desert as observed for solar-like stars,
and of a scaled version of it for lower primary masses (BDs/VLMSs), and
for higher primary masses (giants).
}
\end{figure}

\begin{figure}[b]
\centering
\includegraphics[height=.7\textwidth,angle=0]{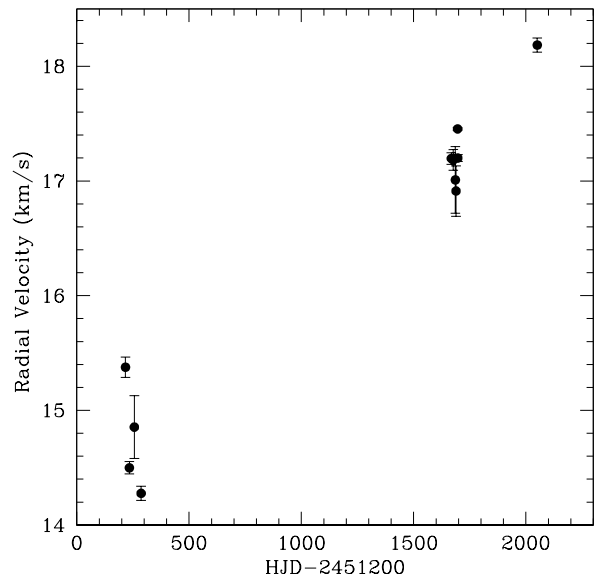}
\caption{
\label{joe:fig3}
Radial velocity data for the low-mass star
CHXR\,74 (M4.5) recorded with UVES/VLT: significant
variability occurring on time scales of months to years hint at a companion
at $a>0.4\,$\,AU and a mass Msin\,$i$ of at least 19 M$_\mathrm{Jup}$
(Joergens 2005 \cite{joe:j2005}).
}
\end{figure}

If BDs form in the same way as stars, we should observe
an equivalent companion mass distribution for both. Then, the BD desert observed for 
solar-like
stars could exist as a scaled-down equivalent also around BD primaries,
this would be a \emph{giant planet desert}, as illustrated in Fig.\,\ref{joe:fig4} 
(Joergens 2005 \cite{joe:j2005}).
The here presented RV survey started to test its existence for BDs in Cha\,I.
For higher than solar-mass primaries,
there are indeed hints that the BD desert might be a scalable phenomenon: 
while RV surveys of K giants detected a much higher rate of 
close BD companions compared to solar-like stars
(e.g. Hatzes et al. 2005 \cite{joe:hatzes}, 
Mitchell et al. 2005 \cite{joe:mitchell}),
with only one exception they all do not lie in the BD
desert when scaled up for the higher primary masses.

\section{Conclusions}

The study of the multiplicity properties of BDs
for separations of less than $\sim$3--10\,AU has been recognized as one of the
main observational efforts that are necessary in order to constrain the formation of BDs.
This can be done by means of high-resolution spectroscopic surveys.
We have presented here the current results of the first RV survey of very young
BDs (Joergens 2005 \cite{joe:j2005}). It exploits the high resolving power and
stability of the UVES spectrograph and the large photon collecting area of the
VLT.
A remarkable feature of this survey is that it is sensitive to
planetary mass companions. This is due to a precise RV
determination and to the fact that systematic RV errors
caused by activity are sufficiently small for the targets
to allow for the detection of Jupiter
mass planets around them, as shown for the first time by this survey.
Thus, very young BDs, at least in Cha\,I, are suitable targets for the search for 
close extrasolar planets in contrast to very young stars.

None of the BDs and VLMSs monitored shows signs of BD or planetary companions
for separations smaller than 0.1\,AU. This hints at a small binary fraction
and a low frequency of giant planets in this separation range (zero of ten).
Within the limited statistics, this result is
consistent with the binary frequency found for field BDs/VLMSs
(12$\pm$7\%, Guenther \& Wuchterl 2003 \cite{joe:gw})
and with the frequency of stellar G dwarf binaries (7\%,
\cite{joe:dm91})
in the same separation range.
For some of the Cha\,I targets also larger separations were probed leading to
the detection of two candidates for spectroscopic systems:
Both the BD candidate Cha\,H$\alpha$\,8 (M6.5) and the low-mass star CHXR74 (M4.5)
exhibit long-term RV variations that were attributed to orbiting companions
with several Jupiter masses at minimum. Orbit solutions have to await
follow-up observations, however, the data suggest orbital periods
of at least several months, i.e. separations of $>$ 0.2\,AU and 0.4\,AU, resp.

Direct imaging surveys found a significantly lower frequency of
BD binaries with separations $a>$3--10\,AU compared
to solar-like stars \cite{joe:bouy}. This might be (partly) caused by
a shift to smaller separations for lower mass primaries.
The first surveys that probe the inner
few AU around BDs by spectroscopic means are the one presented here (Joergens \& Guenther 2001 \cite{joe:j2001},
Joergens 2005 \cite{joe:j2005}, fair detection efficiency for $a<$0.4\,AU \cite{joe:maxted}, sensitive to
M$_{\mathrm{Jup}}$ planets), and the following ones by other groups:
Basri \& Mart\'{\i}n (1999 \cite{joe:basri}, detection of first spectroscopic BD binary),
Reid et al. (2002 \cite{joe:reid}, single epoch spectra, sensitive to
double-lined spectroscopic binaries),
Guenther \& Wuchterl (2003 \cite{joe:gw}, fair detection efficiency for $a<$0.7\,AU \cite{joe:maxted},
sensitive to planetary masses) and
Kenyon et al. (2005 \cite{joe:kenyon}, fair detection efficiency for $a<$0.02\,AU \cite{joe:maxted}, sensitive to BD companions).
These surveys do not hint at a higher BD binary fraction at $a<$1\,AU compared to stellar binaries
indicating that also the overall
binary frequency is lower in the substellar than in the stellar regime.
While corrections have to be applied to the observed values
because of selection biases (e.g.
Burgasser et al. 2003)
and sparse sampling of the
velocity
data (e.g. Maxted \& Jeffries 2005 \cite{joe:maxted}),
a primary goal is to enlarge and improve the available data set for
BD spectroscopic binary studies in terms of 
sample sizes, phase coverage, and precision of RV data.

%
%
%



\printindex
\end{document}